\documentclass[a4paper,dvips,english]{revtex4}
\usepackage{mathptmx}
\usepackage[T1]{fontenc}
\usepackage[latin9]{inputenc}
\usepackage{textcomp}
\usepackage{amsmath}
\usepackage{color}
\usepackage{graphicx}
\usepackage{amssymb}
\usepackage{a4p}

\providecommand{\tabularnewline}{\\}

\usepackage{axodraw}

\usepackage{babel}
\makeatother

\begin{document}
\begin{flushright}
ATL-PHYS-PUB-2007-008 \& SN-ATLAS-2007-066
\par\end{flushright}

\title{The $E_{6}$ inspired isosinglet quark and the Higgs boson }

\date{\today }

\author{S. Sultansoy}

\address{\textcolor{black}{Institute of Physics, Academy of Sciences, Baku,
Azerbaijan }\\
\textcolor{black}{and TOBB University of Economics and Technology,
Physics Department, Ankara, Turkey.}}

\author{G. Unel}

\address{University of California at Irvine, Physics Department, Irvine, USA\\
 and CERN, Physics Department, Geneva, Switzerland.}

\begin{abstract}
We consider the experimental implications of the down type isosinglet
quark, $D$, predicted by the $E_{6}$, group to Higgs boson searches
at the LHC. The pair production of $D$ quarks at the LHC and their
subsequent decays $D\rightarrow h\, d$ and $D\rightarrow W\, u$
has been analyzed. For a light Higgs boson of mass O(120 GeV), an
analysis based on fast simulation of the ATLAS detector response shows
that the $b\overline{b}$ channel becomes as efficient as the $\gamma\gamma$
channel for discovering the Higgs particle if $m_{D}<630$ GeV.
\end{abstract}
\maketitle

\section{Introduction}

Although the standard model (SM) explains the results of the particle
physics experiments performed up to now with an accuracy high enough
to withstand even the challenges set by the precision measurements,
it still leaves some of the basic questions open. The number of elementary
particles, the reason for their mass hierarchy, and the unification
of gravity with the other known forces are some examples of these
open issues. The grand unified theories (GUTs) aim to answer at least
some of these questions by imposing a fundamental symmetry between
all known fermions of the same family. This symmetry, manifesting
itself at high energies, is expected to reduce the number of free
parameters in the SM. The experimental implication of extending the
existing $SU_{C}(3)\times SU_{W}(2)\times U_{Y}(1)$ group structure
of the SM to a single gauge group with a large fundamental representation
is the prediction of new particles. The exceptional Lie group $E_{6}$
has been long considered as one of the favorite candidates for such
a GUT gauge symmetry group \citep{R-e6,R-hewet-rizzo}. The new colored
particles predicted by $E_{6}$ are isosinglet quarks, leptoquarks
and diquarks depending on model variations. Some of these particles
may soon be accessible by the LHC experiments \citep{R-atlas-tdr,R-CMS-tdr}.
The current limit on the mass of the down type isosinglet quark is
$m_{D}>199$GeV \citep{PDG}.The upgraded Tevatron could reach $m_{D}\sim300$
GeV \citep{Rosner}, whereas the LHC will cover the region up to $m_{D}\sim1$
TeV for the pair production channel \citep{details} independent of
the mixing between $d$ and $D$ quarks. For the single production
channel, if the sine of the mixing angle exceeds 0.025, the discovery
reach would be $m_{D}\sim1.5$ TeV \citep{single-prod}. In this note,
we consider the possible impact of the mixing between the $E_{6}$
isosinglet quark and the SM down type quarks on the light Higgs boson
$h$ (e.g. $m_{h}<$135 GeV) searches with the ATLAS detector \citep{R-atlas-tdr}.
According to the ATLAS technical design report \citep{R-atlas-tdr},
for such a light Higgs boson, the $h\rightarrow\gamma\gamma$ channel
is the only one that would give a possible discovery with more than
5 $\sigma$ significance with 100 fb$^{-1}$ of integrated luminosity.
However, if the $E_{6}$ model is true and the mass of the lightest
additional quark is suitable, its decays involving the Higgs boson
enhance the Higgs discovery potential. In particular the channel studied
in this note, $p\, p\to D\bar{D}\to h\, j\, W\, j\to b\bar{b}j\,\ell\nu j$,
becomes as efficient as the $\gamma\gamma$ channel for discovering
the Higgs particle if $m_{D}<630$ GeV.

This paper is organized as follows: Section \ref{sec:The-Higgs-interaction}
contains the description of the model and the new interactions involving
the Higgs particle. Section \ref{sec:The-experimental-implications}
briefly discusses the outcomes of the new interaction: A) the possibility
of differentiation between the quarks of this model and any other
model with an additional down type quark and B) an increase in the
Higgs production rate at the LHC. Section \ref{sec:Detection-Technique}
elaborates on the second outcome by describing the performed Monte
Carlo (MC) studies based on a fast simulation of the ATLAS detector
response. Section \ref{sec:Conclusions} summarizes the statistical
analysis that estimates the experimental reach to discover a light
Higgs for the early days of the LHC.

\section{The Higgs interaction\label{sec:The-Higgs-interaction}}

If the SM $SU_{C}(3)\times SU_{W}(2)\times U_{Y}(1)$ group structure
originates from the breaking of the $E_{6}$ GUT scale down to the
electroweak scale, then the extended quark sector should be written
as:

\begin{equation}
\left(\begin{array}{c}
u_{L}\\
d_{L}\end{array}\right),u_{R},d_{R},D_{L},D_{R}\,;\quad\left(\begin{array}{c}
c_{L}\\
s_{L}\end{array}\right),c_{R},s_{R},S_{L},S_{R}\,;\quad\left(\begin{array}{c}
t_{L}\\
b_{L}\end{array}\right),t_{R},b_{R},B_{L},B_{R}\quad.\label{quarks}\end{equation}
As shown, each SM family is extended by the addition of an isosinglet
quark, respectively denoted by the letters $D$, $S$, and $B$. The
mixing between the new and the SM down type quarks is responsible
for the decays of the former. In this study, the intra-family mixing
of the new quarks is assumed to be dominant compared to their inter-family
mixing. Considering a similar mass hierarchy between the SM and the
new quarks, the $D$ ($B$) quark is assumed to be the lightest (heaviest)
one. The Lagrangian relevant for the $D$ quark weak interactions
is :

\begin{eqnarray}
L_{D} & = & \frac{\sqrt{4\pi\alpha_{em}}}{2\sqrt{2}\sin\theta_{W}}\left[\bar{u}^{\theta}\gamma_{\alpha}\left(1-\gamma_{5}\right)d\cos\phi+\bar{u}^{\theta}\gamma_{\alpha}\left(1-\gamma_{5}\right)D\sin\phi\right]W^{\alpha}\label{lagrangian}\\
 & - & \frac{\sqrt{4\pi\alpha_{em}}}{4\sin\theta_{W}}\left[\frac{\sin\phi\cos\phi}{\cos\theta_{W}}\bar{d}\gamma_{\alpha}\left(1-\gamma_{5}\right)D\right]Z^{\alpha}\nonumber \\
 & - & \frac{\sqrt{4\pi\alpha_{em}}}{12\cos\theta_{W}\sin\theta_{W}}\left[\bar{D}\gamma_{\alpha}\left(4\sin^{2}\theta_{W}-3\sin^{2}\phi(1-\gamma_{5})\right)D+\bar{d}\gamma_{\alpha}\left(4\sin^{2}\theta_{W}-3\cos^{2}\phi(1-\gamma_{5})\right)d\right]Z^{\alpha}\;+h.c.\nonumber \end{eqnarray}

where $\phi$ is the mixing angle between the $d$ and $D$ quarks,
and $\theta$ represents the usual CKM quark mixings, taken to be
in the up sector for simplicity of calculation. Assuming that the
squared sum of the row elements of the new 3 \texttimes{} 4 CKM matrix
gives unity, the measured values of $V_{ud}$,$V_{us}$, and $V_{ub}$
\citep{PDG} constrain $\sin\phi$ $\leq0.07$~\citep{details}.
The total decay width and the contribution from neutral and charged
currents were already estimated in \citep{R-e6-orhan-metin}. As reported
in that work, the $D$ quark decays through a $W$ boson with a branching
ratio of 67\% and through a $Z$ boson with a branching ratio of 33\%. 

The origin of the masses of SM particles is explained by using the
Higgs mechanism. The Higgs mechanism can also be preserved in $E_{6}$
group structure as an effective theory, although other alternatives
such as dynamical symmetry breaking are also proposed \citep{DSB}.
On the other hand, the origin of the mass of the new quarks ($D,\, S,\, B$)
should be due to another mechanism since these are isosinglets. Therefore,
the mass terms in the Lagrangian, written in the mass basis, are simply:\begin{equation}
L_{dD}^{M}=m_{d}\bar{d}_{L}^{M}d_{R}^{M}+m_{D}\bar{D}_{L}^{M}D_{R}^{M}+h.c.\label{eq:mass-basis}\end{equation}

where $m_{d}$ and $m_{D}$ are the masses of the $d$ and $D$ quarks
respectively. Based on this initial consideration, the studies in
\citep{R-e6-orhan-metin,details} did not consider the channels involving
the Higgs boson $h$ for the $D$ quark decays. However, the mixing
between $d$ and $D$ quarks will lead to decays of the latter involving
$h$ after spontaneous symmetry breaking (SSB). To find these decay
channels, the interaction between the Higgs field and both down type
quarks of the first family should be considered before SSB. In this
study, the Lagrangian in SM basis, containing also the cross terms
between $d$ and $D$ with mass-like coefficients ($a_{dD}$, $m_{Dd}$),
is written as: 

\begin{eqnarray}
L_{dD}^{0} & = & a_{dd}\bar{d}_{L}^{0}d_{R}^{0}H+a_{dD}\bar{d}_{L}^{0}D_{R}^{0}H\label{eq:main}\\
 & + & m_{Dd}\bar{D}_{L}^{0}d_{R}^{0}+m_{DD}\bar{D}_{L}^{0}D_{R}^{0}+h.c.\qquad.\nonumber \end{eqnarray}

\begin{figure}
\begin{centering}
\includegraphics[scale=0.4]{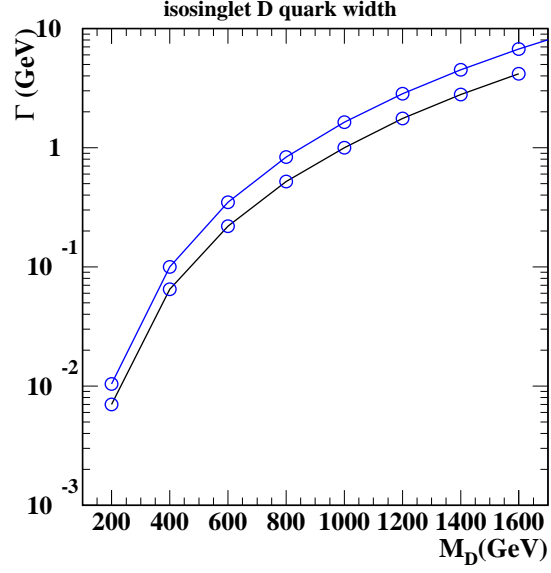}
\par\end{centering}

\caption{The decay width of the $D$ quark for $\sin\phi=0.045$: The lower
curve corresponds to the case without Higgs interaction, whereas the
upper curve has been calculated by taking into account the Higgs channels
for $m_{H}=120$ GeV .\label{fig:The-decay-width}}

\end{figure}

\begin{figure}
\begin{centering}
\includegraphics[scale=0.4]{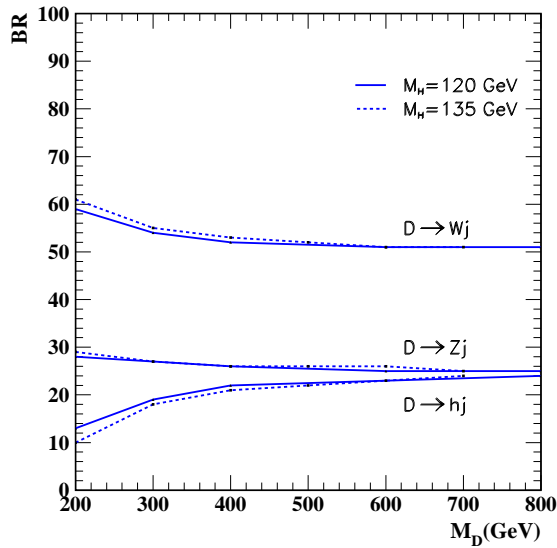}
\par\end{centering}

\caption{The D quark branching ratios as a function of the D quark mass for
two Higgs mass values: 120 GeV \& 135 GeV.\label{fig:The-D-quark-BR}}

\end{figure}

The four interaction coefficients ($a_{dd},\, a_{dD},\, m_{Dd},\, m_{DD}$)
in equation (\ref{eq:main}) can be obtained in terms of the two quark
masses and the two mixing angles between the left and right components
of the two down type quarks. In order to use the new variable set,
the first step is to write the mixing between the SM and the mass
states as follows:

\begin{eqnarray}
D_{L/R}^{0} & = & D_{L/R}^{M}\cos\phi_{L/R}-d_{L/R}^{M}\sin\phi_{L/R}\label{eq:solving}\\
d_{L/R}^{0} & = & D_{L/R}^{M}\sin\phi_{L/R}+d_{L/R}^{M}\cos\phi_{L/R}\nonumber \end{eqnarray}
where $\phi_{L/R}$ is the $d$-$D$ quark mixing angle with the subscript
L (R) standing for the left (right) components of the quark fields
and the superscript 0 (M) standing for the SM (mass) basis. After
the SSB, the Higgs field is expanded around the new minimum, $\nu$,
as $H=\nu+h$ to separate the terms that contribute to interaction
and mass coefficients:

\begin{eqnarray}
\textrm{$a_{dd}$} & = & \frac{m_{d}\cos\phi_{L}\cos\phi_{R}+m_{D}\sin\phi_{L}\sin\phi_{R}}{\nu}\label{eq:solution}\\
a_{dD} & = & \frac{m_{d}\cos\phi_{L}\sin\phi_{R}-m_{D}\sin\phi_{L}\cos\phi_{R}}{\nu}\nonumber \\
m_{Dd} & = & m_{d}\sin\phi_{L}\cos\phi_{R}-m_{D}\cos\phi_{L}\sin\phi_{R}\nonumber \\
m_{DD} & = & m_{d}\sin\phi_{L}\sin\phi_{R}+m_{D}\cos\phi_{L}\cos\phi_{R}\nonumber \end{eqnarray}

where $\nu=\eta/\sqrt{2}$ and $\eta=247$ GeV is the vacuum expectation
value of the Higgs field. After simple calculations, the Lagrangian
for the interaction between $d,\; D$ quarks, and the Higgs boson
becomes : 

\begin{eqnarray}
L_{h}^{M} & = & \frac{m_{D}}{\nu}\sin^{2}\phi_{L}\bar{D}^{M}D^{M}h\label{eq:DDh}\\
 & - & \frac{{\sin{\phi_{L}}\cos{\phi_{L}}}}{2\nu}\bar{{D}}^{M}\left[(1-\gamma^{5})\, m_{D}+(1+\gamma^{5})\, m_{d}\right]\, d^{M}\, h\nonumber \\
 & - & \frac{{\sin{\phi_{L}}\cos{\phi_{L}}}}{2\nu}\bar{{d}}^{M}\left[(1+\gamma^{5})\, m_{D}+(1-\gamma^{5})\, m_{d}\right]\, D^{M}\, h\nonumber \\
 & + & \frac{m_{d}}{\nu}\cos^{2}\phi_{L}\bar{d}^{M}d^{M}h\nonumber \end{eqnarray}
Note that the right mixing angle has completely disappeared from the
final formula (\ref{eq:DDh}). The total width of the $D$ quark as
a function of its mass is shown in Fig.~\ref{fig:The-decay-width}
for the illustrative value of $\sin\phi=0.045$. It is seen that the
$D$~quark has a narrow width and becomes even narrower with decreasing
values of $\phi$ since it scales through a $\sin^{2}\phi$ dependence.
The relative branching ratios for the decay of the $D$~quark depend
on both the $D$~quark and the Higgs mass values. For example, at
the values of $D$~quark mass around 200~GeV and the Higgs mass
around 120$\;$GeV: Br($D\rightarrow Wu$)$\sim$60\%, Br($D\rightarrow hd$)$\sim$12\%,
Br($D\rightarrow Zd$)$\sim$28\%, whereas as the $D$~quark mass
increases the same ratios asymptotically reach 50\%, 25\% and 25\%
respectively. As the Higgs mass increases from 120~GeV, these limit
values are reached at higher $D$~quark masses. The branching ratios
as a function of the $D$~quark mass are given in Fig.~\ref{fig:The-D-quark-BR}
for two values of the Higgs mass: 120~GeV (solid lines), which is
just above the limit imposed by LEP II results, and 135~GeV (dashed
lines), where the $b\bar{b}$ decay mode is taken over by the $W^{+}W^{-}$mode.

\section{The experimental implications\label{sec:The-experimental-implications}}

Studies about the $D$~quark pair production without the Higgs particle
have been previously reported elsewhere \citep{details}. This section
concentrates on the possible impact of the interactions involving
the Higgs particle. For example, equation (\ref{eq:DDh}) shows that,
the Higgs boson could be used in the $s$ channel for the creation
of $D-d$ quark pairs: $h\to\bar{D}\, d$ or $\bar{d}\, D$. However,
as the Higgs couplings depend on the quark masses, the right handed
$D$ quarks ($D_{R}$) would be produced $(\frac{m_{D}}{m_{d}})^{2}$
times more often than the left handed ones ($D_{L}$). Due to the
left handedness of the charged weak interactions, the produced $D$
quarks would subsequently decay mostly via the neutral current channel.
The precise measurement of such a process could help in differentiating
the $E_{6}$ model from other models that also involve additional
down type quarks. 

\begin{figure}
\begin{centering}
\includegraphics[scale=0.4]{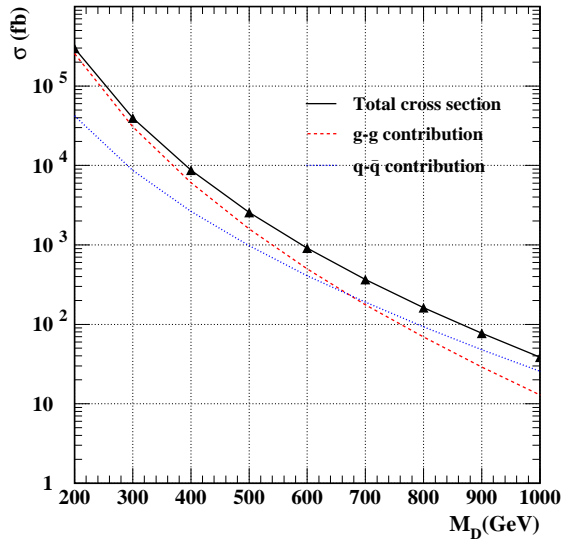}
\par\end{centering}

\caption{Tree level pair production of $D$ quarks at LHC as a function of
the new quark's mass. \label{fig:D-Pair-prod}}

\end{figure}

Depending on the masses of the $D$ quark and the Higgs boson itself,
the $E_{6}$ model could also boost the overall Higgs production at
the LHC. This boost is particularly interesting for the Higgs hunt,
one of the main goals of the ATLAS experiment. For example, if the
$D$~quark mass is as low as 250~GeV, its pair production cross
section at the LHC becomes as high as $10^{5}$ fb as can be seen
in Fig.~\ref{fig:D-Pair-prod}. Convolution of the cross section
with the $D$~quark decay branching ratios from previous section
(17\% for $D\to h\, j$ at m$_{D}$=250 GeV) yields the number of
expected Higgs events. In the low mass range considered in this note
(from 115 up to 135~GeV), the branching ratio $h\rightarrow b\overline{b}$
is about 70\% \citep{R-atlas-tdr}. The signatures of the expected
final states involving at least one Higgs boson are listed in table
\ref{tab:expected-final-states} together with the yearly expected
number of events for two example $D$ quark masses: 250 and 500 GeV.
For the remaining of this note, although the case involving the $Z$
is more suitable from the event reconstruction point of view, we will
concentrate on the last row, which has the highest number of expected
Higgs events per year. 

\begin{table}
\caption{For pair production of D quarks, the decay channels involving at least
one Higgs boson. The branching ratios and the number of expected Higgs
particles are calculated assuming $m_{H}$=120 GeV and $m_{D}$=250
(500) GeV. \label{tab:expected-final-states}}

\begin{tabular}{c|c|c|c|c}
$D_{1}$ & $D_{2}$ & BR & \#expected Higgs/100fb$^{-1}$ & expected final state\tabularnewline
\hline 
$D\rightarrow h\, j$ & $D\rightarrow h\, j$ & 0.029 (0.053) & 0.58$\times10^{6}$ (2.65$\times10^{4}$ ) & $2j\;4j_{b}$\tabularnewline
\hline 
$D\rightarrow h\, j$ & $D\rightarrow Z\, j$ & 0.092 (0.120)  & 0.92$\times10^{6}$ (3.01$\times10^{4}$) & $2j\;2j_{b}\;2l$\tabularnewline
\hline 
$D\rightarrow h\, j$ & $D\rightarrow W\, j$ & 0.190 (0.235) & 1.9$\times10^{6}$ (6.04$\times10^{4}$) & $2j\;2j_{b}\; l\; E_{T,miss}$\tabularnewline
\end{tabular}
\end{table}

\section{Monte Carlo Study\label{sec:Detection-Technique}}

The Lagrangian presented in section \ref{sec:The-Higgs-interaction}
has been implemented in a tree level event generator, Comphep v4.4.3.\citep{R-calchep},
to investigate the possibility of detecting the Higgs particle and
reconstructing it from $b-$jets using the ATLAS detector. Assuming
a light Higgs boson of mass 120 GeV, four mass values for the $D$~quark
have been taken as examples: 250~GeV, 500~GeV, 750~GeV, and 1000~GeV.
10$\,$000 signal events were produced for each mass value under study
with the $W\, h\, j\, j\quad(h\to b\bar{b}$ \& $W\to\ell\nu)$ final
states using the CTEQ6L1 PDF set \citep{R-cteq}. Other parameterizations,
such as CTEQ5M and CTEQ5L1 were also tested to investigate the systematic
effect of the PDF selection on the $D\overline{D}$ pair production
total cross section. It was found that the difference was less than
10\% over the range of 200 to 1600 GeV, CTEQ6L1 giving the lowest
results up to a $D$~quark mass of 1 TeV. The generator level cuts
on the partons, guided by the performance of the ATLAS detector, are
listed as:

\begin{eqnarray*}
|\eta_{p}| & \leq & 3.2\quad,\\
p_{T\, p} & \geq & 15\,\hbox{GeV}\quad,\\
R_{p} & > & 0.4\end{eqnarray*}

where $\eta_{p}$ is the pseudo-rapidity for the partons giving rise
to jets; $p_{T\, p}$ is the transverse momentum of the partons; and
$R_{p}$ is the angular separation between the partons. The imposed
maximum value of $\eta_{p}$ requires the jets seeded by the partons
to be in the extended barrel region of the calorimeter where the jet
energy resolution is optimal. The imposed lower value of $p_{T\, p}$
ensures that no jets that would eventually go undetected along the
beam pipe are generated at all. The imposed lower value of $R_{p}$
provides good separation between the two jets in the final state.
Using the interface provided by CPYTH v2.3 \citep{Cpyth}, the generated
particles are processed within the ATLAS software framework version
11.0.41. The decay and hadronization processes were handled in PYTHIA
\citep{Pythia} and detector response simulation was performed with
fast Monte Carlo program ATLFAST \citep{ATLFast} which uses a general
parameterization of the detector, ignoring the individual event topology.
The default properties of the parameterized fast simulation are discussed
elsewhere \citep{details,ATLFast}. One should also note that in this
work, the reconstructed $b-$jet energy and momenta were re-calibrated
like in \citep{R-atlas-tdr} to have a good match between the mean
value of the reconstructed Higgs mass and its parton level value. 

As for the background study, all the SM interactions giving the $W^{\pm}j_{b}\, j_{b}\, j\, j$
final state have been computed in another tree level generator, MadGraph
v2.1. \citep{MadGraph}, using the same parton level cuts and PDF.
The SM background cross section is calculated to be 520 $\pm$11pb.
The reasons for using two separate event generators, their compatibility,
and their relative merits have been discussed elsewhere \citep{details}.
The generated 40$\,$000 background events were also processed in
the same way using the ATLAS software framework for hadronization
and calculation of detector effects.

\subsection{Analysis Details for $m_{D}=500$ GeV}

\begin{figure}
\begin{centering}
\includegraphics[scale=0.7]{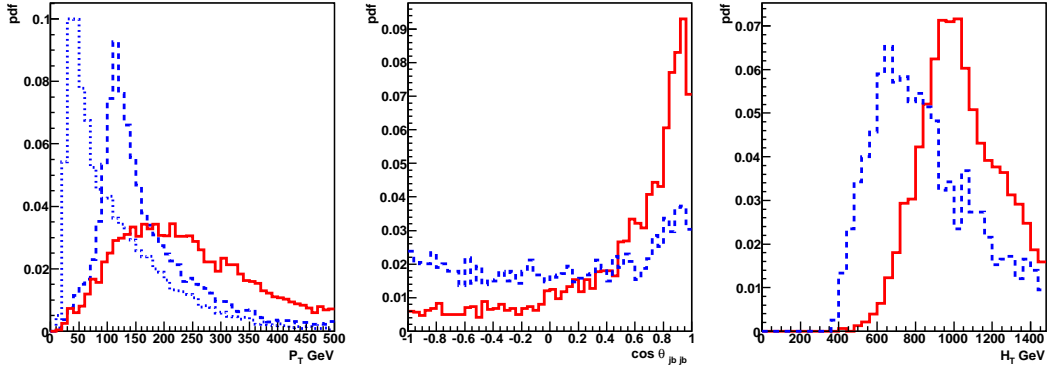}
\par\end{centering}

\caption{The distributions of kinematic parameters for SM background and signal
for an example $D$ quark mass of value of 500 GeV. The solid histograms
are for signal and the dashed ones for the background events. The
leftmost plot contains the $p_{T}$ distribution for the light jets
which show clear distinction between signal and background cases.
It also contains the $p_{T}$ distribution for for the $b$-jets (dotted
histogram) which has the same shape for both signal and background.
However, the middle plot shows that the $b$-jets originating from
the signal events can be discriminated from the background ones by
considering the angle between the two jets. Finally the rightmost
plot contains the scalar sum of all transverse momenta providing another
criterion for separating the signal from the background. \label{fig:Kinematic-variable-distributions}}

\end{figure}

The $W$ boson will be reconstructed using an electron or a muon and
the missing energy associated with the undetected neutrino, whereas
the $b-$jets will be combined to reconstitute the $h$ boson. These
two bosons can be merged together with the remaining two jets to obtain
the invariant masses of the two $D$ quarks. Therefore, the final
state particles of interest are $j_{b}\, j_{b}\, j\, j\,\ell\, E_{T,miss}$
for which the distributions of the kinematic values are shown in Fig.~\ref{fig:Kinematic-variable-distributions}.
On the leftmost plot, one can observe that the $p_{T}$ distribution
of the jets coming from $D$ quark decays peaks around 200 GeV (solid
line), whereas the same distribution for the background jets is much
less energetic: the $p_{T}$ distribution of the $b-$jets peaks around
40 GeV (dotted line) and for other jets the peak is around 120 GeV
(dashed line). This difference can be used to reject the background
events most of which (about 80\%) are originating from the $t\,\bar{t}$
production. In these events, one $W$ decaying leptonically and the
other hadronically give rise to the same final state particles as
the signal. Another property that can be used to discriminate the
background events is the fact that two $b-$jets in the signal originate
from the same particle, whereas for most of the background cases,
they originate from two different particles. The cosine of the angle
between these jets, shown in Fig.~\ref{fig:Kinematic-variable-distributions}
middle plot, is therefore peaked at one for the signal events, as
opposed to the background case, where the distribution is more uniform.
The scalar sum of all transverse momenta in an event ($H_{T}$), is
given in the same figure, rightmost plot, with again a solid line
for the signal events and with a dashed line for the background events.
This variable will be of great benefit for distinguishing the signal
events from the background, especially for the high $D$ quark mass
values.

The reconstruction starts with the requirement of 1 electron or 1
muon and at least 4 jets, two of them identified as $b-$jets. For
the leptons, the transverse momentum minimum value ($p_{T,\ell}>15$
GeV) is selected. The transverse momentum requirement on the two most
energetic jets and the cosine of the angle between the two $b-$jet
are optimized to get the best signal to background ratio for the final
$D$ quark invariant mass distribution. The $h$ boson invariant mass
is formed using the two $b$ tagged jets remaining after the angular
cut. To further reduce the background, the invariant mass of the two
non $b$ tagged jets is also calculated, and the events yielding a
value smaller than $M_{Z}$ are rejected. As there are no $Z$ bosons
in this search channel, this cut ensures the purity of the selected
events by removing the background events originating from the hadronic
decays of the SM $Z$ boson. Another rejection criteria applied at
this stage is related to $H_{T}$, where only events with $H_{T}>800$
GeV are selected. The association between the two most energetic jets
and the reconstructed bosons is not unique. Both possibilities are
calculated and the case with the smallest difference between the two
reconstructed $D$ quark invariant masses is selected. As the last
consistency check, the difference between the two reconstructed $D$
quark masses is required to be smaller than 100 GeV, a value compatible
with the resolution of the reconstructed mass. The event selection
cuts imposed on both signal and background are given in table \ref{tab:Event-selection-cuts}
in the order they were applied in the analysis. The same table also
contains the individual efficiency values of each cut for both signal
and background cases with an error of a few percent. As can be calculated
from the table, the final efficiency for the signal is about 11\%
and for the background is about 2\%.

\begin{table}
\caption{Optimized event selection cuts and their efficiencies for $m_{D}=500$
GeV.\label{tab:Event-selection-cuts}}

\begin{tabular}{c|c|c|c|c|c|c|c|c|c}
 & N-leptons & N-jets & N-bjets & $p_{T}-$lepton & $p_{T}\,$ jet & $\cos\theta_{j_{b}\, j_{b}}$ & $M_{j\, j}$ & $H_{T}$ & $|m_{D1}-m_{D2}|$\tabularnewline
\hline 
Cut value & =1 & $\geq4$ & $\geq2$ & $\geq15$ GeV & $\geq100$GeV & $\geq$-0.8 & $\geq90$GeV & $\geq800$ GeV & $\leq100$GeV\tabularnewline
\hline 
$\epsilon_{signal}$(\%) & 83 & 99 & 33 & 95 & 83 & 97 & 99 & 90 & 59\tabularnewline
\hline
$\epsilon_{backgr}$(\%) & 79 & 99 & 36 & 94 & 69 & 89 & 65 & 55 & 37\tabularnewline
\end{tabular}
\end{table}

\begin{figure}
\begin{centering}
\includegraphics[scale=0.6]{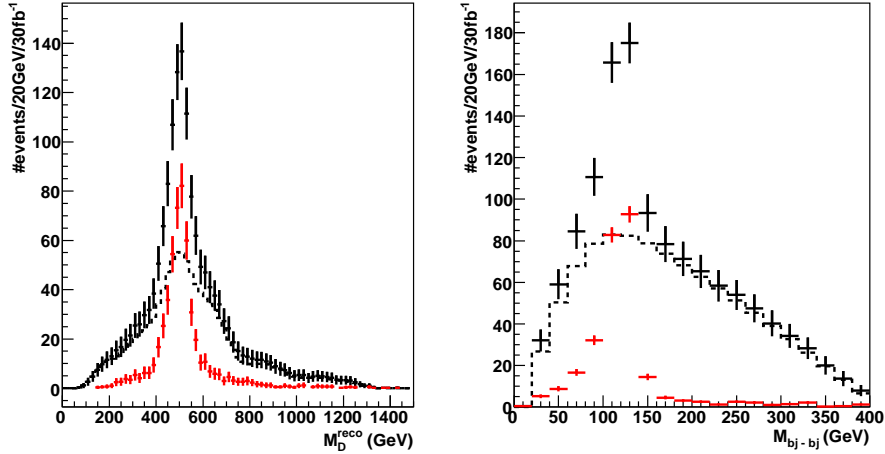}
\par\end{centering}

\caption{Reconstructed invariant masses of the $D$ quark (left) and of the
Higgs boson (right) together with the SM background (dotted lines)
after 30 fb$^{-1}$ integrated luminosity. The mass of the $D$ quark
is set to 500 GeV and Higgs boson to 120 GeV.}

\label{fig:inv-mass-DH}
\end{figure}

The invariant mass distributions after all cuts for the $D$ quark
and the Higgs boson are presented in Fig.~\ref{fig:inv-mass-DH}
for 30 fb$^{-1}$ integrated luminosity. After defining the signal
region for $D$ as $M_{D}\pm50$ GeV and for $h$ as $M_{h}\pm30$
GeV, the number of events for the signal (S) and the background (B)
can be integrated for both to obtain the statistical significance
$\sigma=S/\sqrt{S+B}$. For this set of parameters, it is found that
the $D$ quark can be observed with a significance of 13.2$\sigma$
and at the same time the Higgs boson with a significance of about
9.5$\sigma$. One should note that, in the SM Higgs searches, such
a high statistical significance can only be reached with more than
3 times more data: with about 100 fb$^{-1}$ integrated luminosity. 

\subsection{Extension to other mass values}

\begin{figure}
\includegraphics[width=0.3\columnwidth]{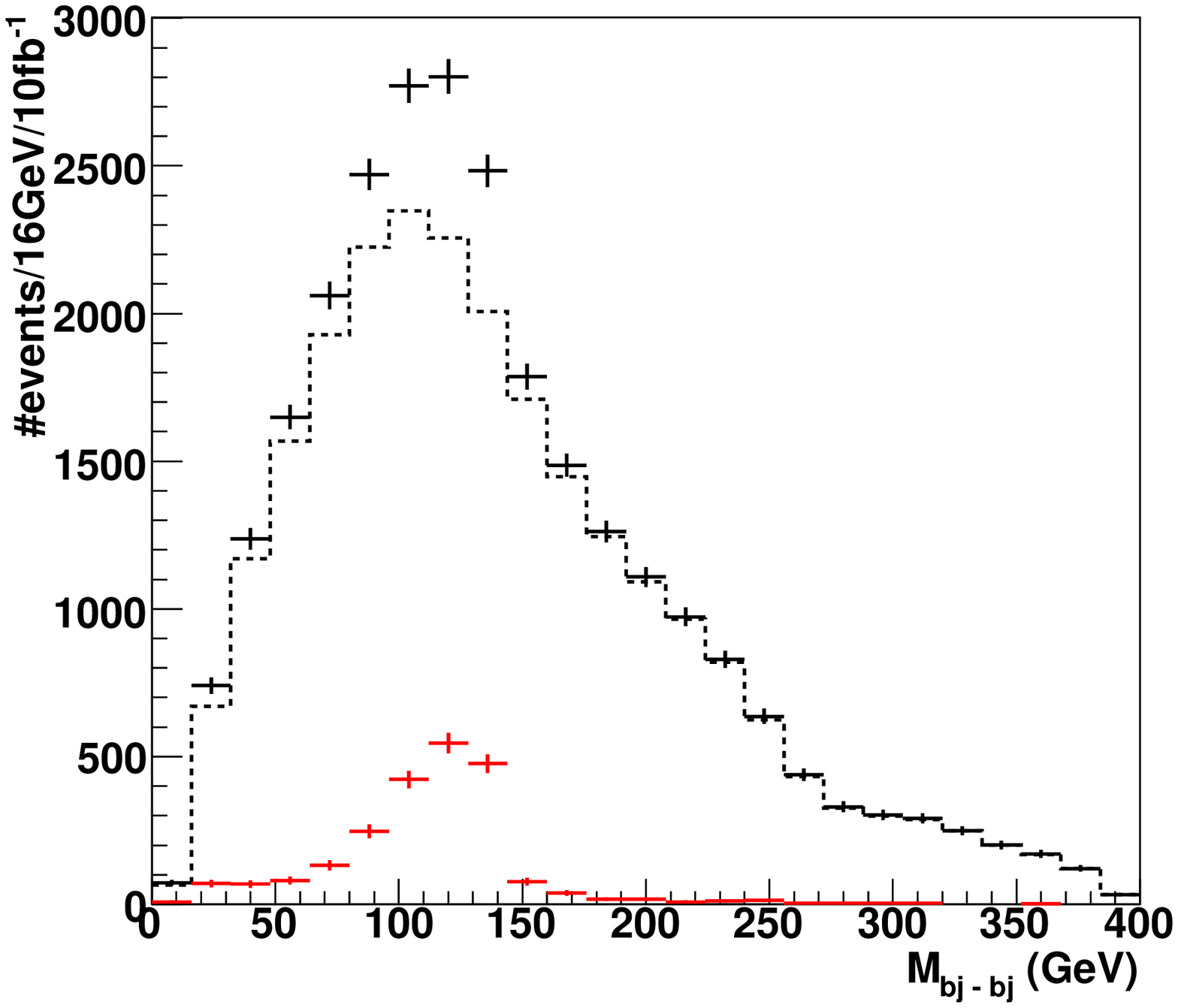}\includegraphics[width=0.3\columnwidth]{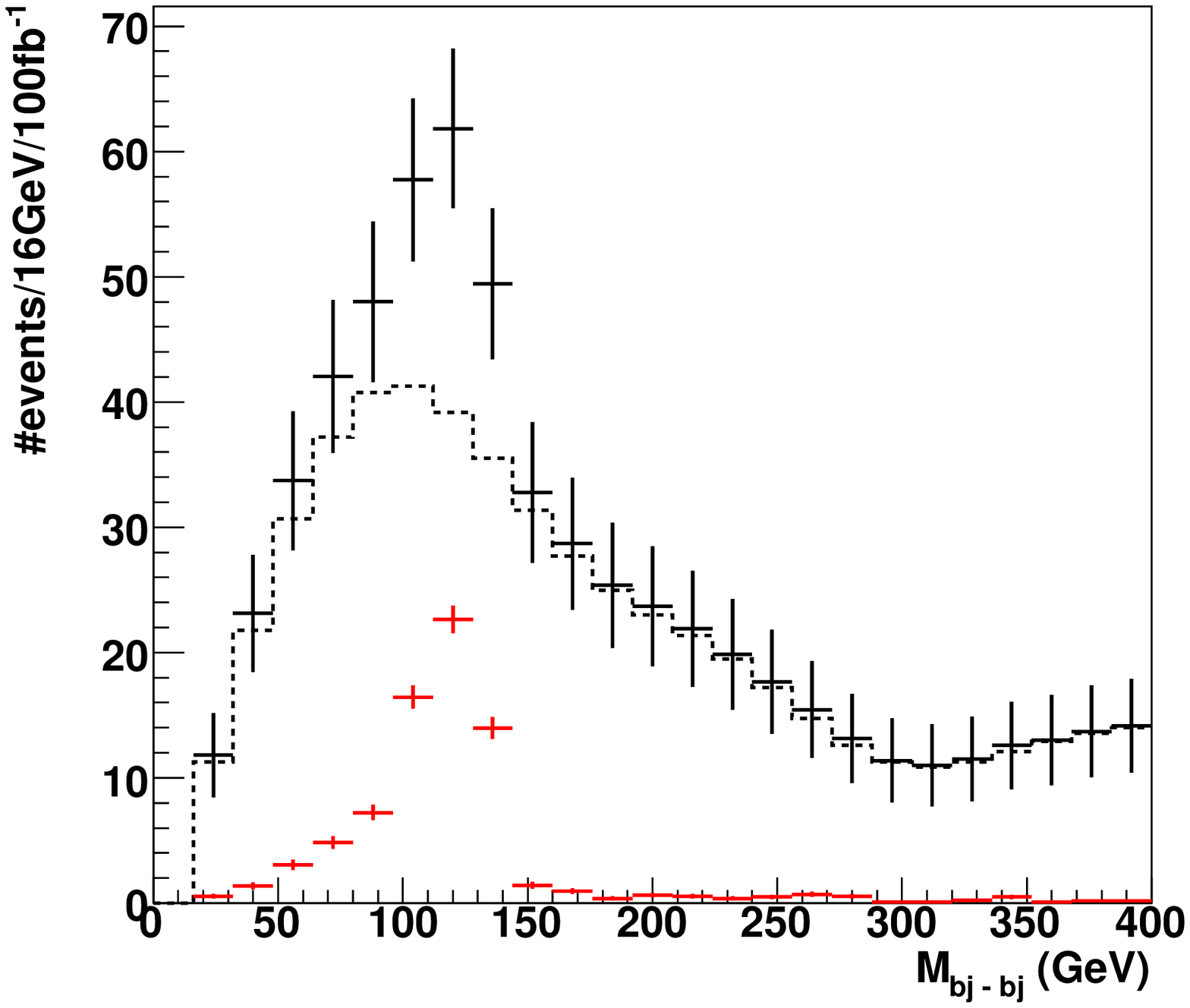}\includegraphics[width=0.3\columnwidth]{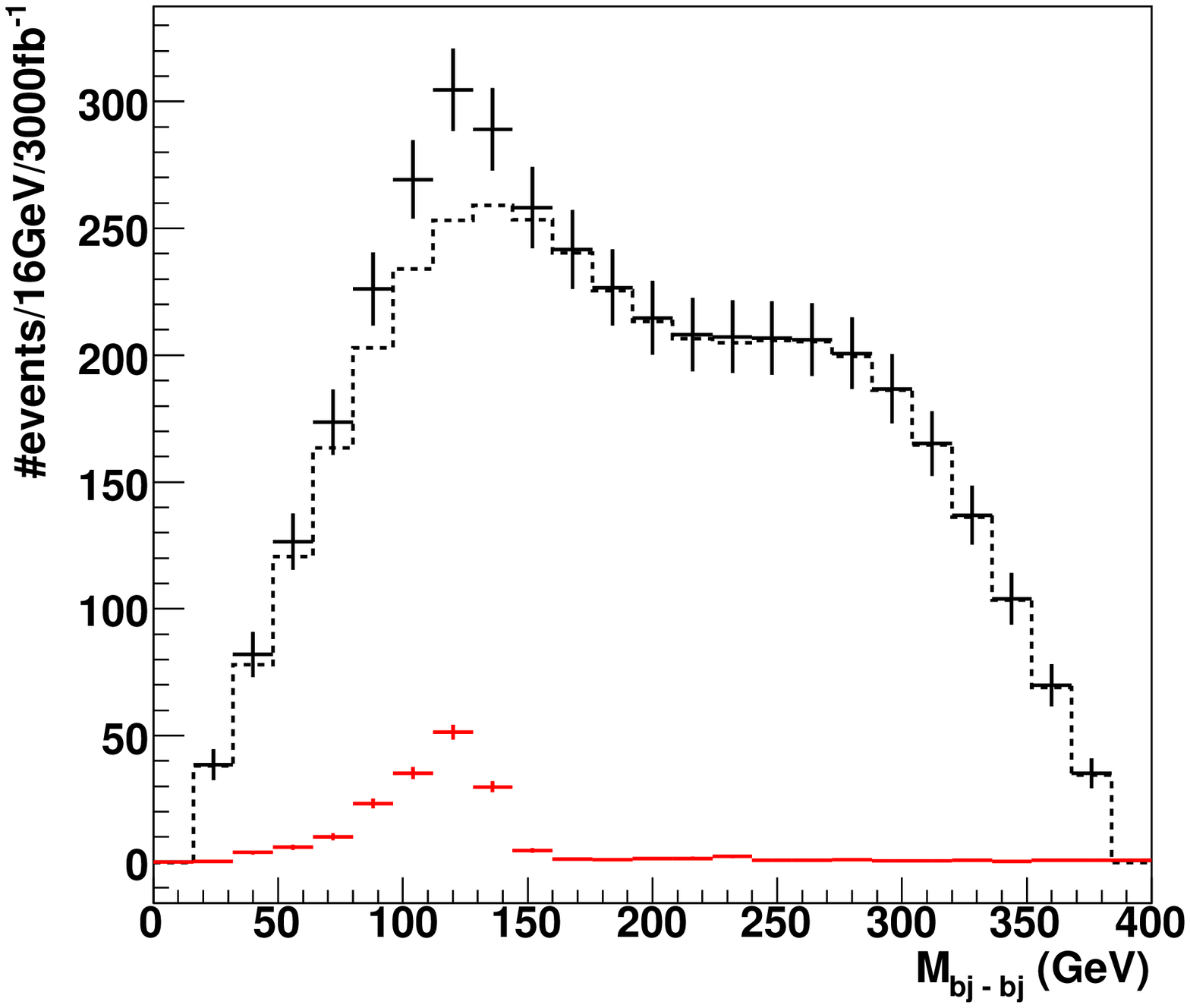}

\caption{The reconstructed masses for the Higgs particles for different $D$
quark mass values. The dashed histogram represents the SM background,
the red data points are for the signal case and the black data points
are the sum of signal and the background. From left to right 250,
750 \& 1000 GeV $D$ quarks were considered using 10, 100 and 3000
fb$^{-1}$ integrated luminosities respectively.}

\label{fig:others}
\end{figure}

An analysis similar to the one presented in the previous section was
performed for the other three $D$ quark masses: 250, 750 and 1000
GeV. For each mass, the cut values were re-optimized to get the best
statistical significance in the Higgs boson search. The method is
to start from a low value of the $H_{T}$ cut and scan the available
sample by increasing the cut value, like a sliding threshold to look
for a peak in the reconstructed mass histogram. The invariant mass
distribution of the reconstructed Higgs boson for different $D$ quark
mass values is given in Fig.~\ref{fig:others} where the dashed line
shows the SM background, the red data points are for the signal events
only and the black data points are the sum of the signal and the background
events. For each increasing $D$ quark mass value under consideration,
the total integrated luminosity was also increased to be able to observe
the Higgs boson signal.

Table \ref{tab:The-expectations} gives the number of events for signal
and background processes after 3 years of low luminosity data taking
time (30 fb$^{-1}$ integrated luminosity) for all considered $D$
quark mass values. With such integrated luminosity, the 3$\sigma$
signal observation limit for the $D$ quark is at 750 GeV. If $m_{D}\geq1000$
GeV, the required luminosities can be attained either by running the
LHC for many years at the design parameters or by considering a possible
upgrade, named as Super-LHC, which would yield 1000 fb$^{-1}$ per
year. On the other hand, the significance values presented in Table
\ref{tab:The-expectations} could be reduced by systematic errors
which have not been evaluated here, and by the dilution of the statistical
significance which occurs in searches for an excess in any bin of
distribution.

\begin{table}
\caption{The expected number of signal (S) and background (B) events for $D$
quark and Higgs boson searches after 30 fb$^{-1}$ integrated luminosity.
The counted events are in a range of $\pm$50 GeV ($\pm$20 GeV) from
the generator level value of the $D$ quark (Higgs boson). }

\begin{centering}
\begin{tabular}{c|c|c|c|c|c|c|c|c}
$M_{D}$(GeV) & \multicolumn{2}{c|}{250} & \multicolumn{2}{c|}{500} & \multicolumn{2}{c|}{750} & \multicolumn{2}{c}{1000}\tabularnewline
\cline{2-3} \cline{4-5} \cline{6-7} \cline{8-9} 
\hline 
$S$ & 8802 & 5303 & 336 & 222 & 27 & 19 & 1.9 & 1.4\tabularnewline
\hline 
$B$  & 29379 & 31717 & 313 & 321 & 32 & 56 & 3.1 & 10.6\tabularnewline
\hline 
$S$/$\sqrt{B+S}$ & 45.1 & 27.6 & 13.2 & 9.5 & \multicolumn{1}{c|}{3.5} & 2.1 & 0.84 & 0.42\tabularnewline
\end{tabular}\label{tab:The-expectations}
\par\end{centering}
\end{table}

\begin{figure}

\begin{centering}
\includegraphics[scale=0.5]{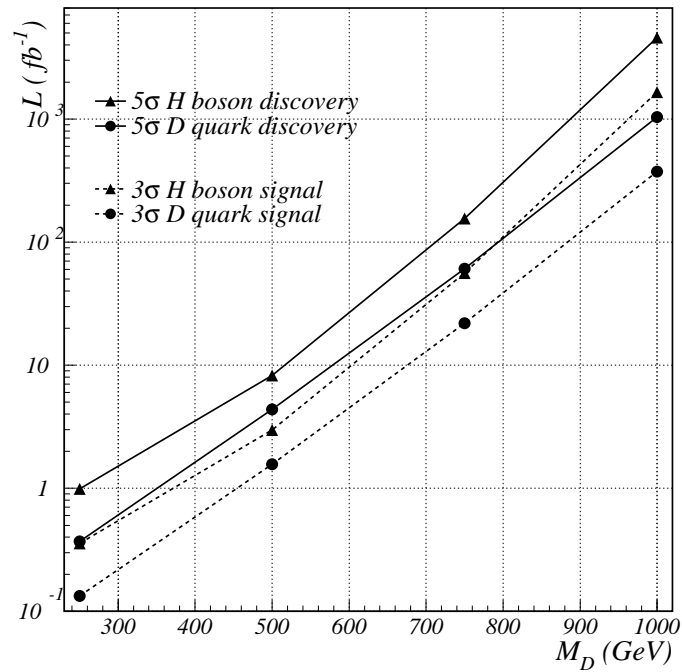}
\par\end{centering}

\caption{The 3 and 5 $\sigma$ reach of ATLAS in the Higgs and $D$ quark search
for increasing $D$ quark mass values}

\label{fig:reach}
\end{figure}

\section{Conclusions\label{sec:Conclusions}}

This work gives the Lagrangian for the $D$ quark including its interaction
with the known SM particles and Higgs boson. One of the two implications
of the Higgs interaction is discussed in detail: the boost to the
Higgs discovery potential at the LHC. It is shown that the ATLAS experiment
can use the $D$ quark decay channels to enhance the Higgs discovery
potential independent of the mixing angle between $d$ and $D$ quarks.
The prospects for the Higgs discovery are discussed using Monte Carlo
and fast simulation techniques for a light Higgs particle (120 GeV)
and some exemplary $D$ quark mass values ranging from 250 to 1000
GeV. Fig.~\ref{fig:reach} contains the 3$\sigma$ (dashed lines)
and the 5 $\sigma$ (solid lines) reaches of Higgs boson (triangles)
and $D$ quark (circles) searches for the above mentioned values.
Therefore, a light Higgs boson could be discovered with a 5$\sigma$
statistical significance using the $D\bar{D}\rightarrow hWjj$ channel
within the first year of low luminosity data taking (integrated luminosity
of 10 fb$^{-1}$) if $m_{D}<500$ GeV. Under the same conditions but
with one year of design luminosity (integrated luminosity of 100 fb$^{-1})$,
the 5 $\sigma$ Higgs discovery can be reached if $m_{D}\leq700$
GeV. This is to be compared with the studies from the ATLAS TDR, where
the most efficient channel to discover such a light Higgs is the $h\rightarrow\gamma\gamma$
decay. This search yields about 8$\sigma$ signal significance with
100 fb$^{-1}$ integrated luminosity. The presently discussed model
could give the same significance (or more) with the same integrated
luminosity if $m_{D}<630$ GeV. Therefore, if the isosinglet quarks
exist and their masses are suitable, they will provide a considerable
improvement for the Higgs discovery potential. The same channel also
provides the possibility to search for the lightest of the isosinglet
quarks, providing a $5\,\sigma$ discovery signal if $m_{D}<800$
GeV, within 100 fb$^{-1}$ of integrated luminosity. However one should
note that the version of the fast simulation software used in this
note was not fully validated by the ATLAS collaboration. The study
of other channels involving leptonic decays of the $Z$ boson with
thoroughly validated versions of the simulation and reconstruction
software is in progress.

\section*{Acknowledgments}

The authors would like to thank Louis Tremblet and CERN Micro Club
for kindly providing computational facilities. We are grateful to
G. Azuelos, J. D. Bjorken, J. L. Rosner, F. Ledroit and A. Parker
for useful discussions and to A. Lankford and B. Golden for useful
suggestions. S.S acknowledges the support from the Turkish State Planning
Committee under the contract DPT2006K-120470. G.U.'s work is supported
in part by U.S. Department of Energy Grant DE FG0291ER40679. This
work has been performed within the ATLAS Collaboration with the help
of the simulation framework and tools that are the results of the
collaboration-wide efforts.


\begin{thebibliography}{10}
\bibitem{R-e6}F. Gursey, P. Ramond and P. Sikivie, Phys. Lett. B
\textbf{60,} \emph{177} (1976); F. Gursey and M. Serdaroglu, Lett.
Nuovo Cimento \textbf{21}, \emph{28} (1978).

\bibitem{R-hewet-rizzo}J. Hewett and T. Rizzo, Phys. Rep. \textbf{183},
\emph{195} (1989). 

\bibitem{R-atlas-tdr}ATLAS Detector and Physics Performance Technical
Design Report. CERN/LHCC/99-14/15.

\bibitem{R-CMS-tdr}CMS collaboration, Technical proposal, CERN-LHCC-94-38.

\bibitem{PDG}Particle Data Group, Phys. Lett. B \textbf{592,} \emph{1}
(2004).

\bibitem{Rosner}T. C. Andre and C.L. Rosner, Phys. Rev. D. \textbf{69},
\emph{035009}, (2004). 

\bibitem{details}R. Mehdiyev et al., ATL-PHYS-PUB-2005-021 (2005);
Euro. Phys. J. C. \textbf{49}, 13 (2007).

\bibitem{single-prod}S. Sultansoy, G. Unel and M. Yilmaz, {[}arXiv:hep-ex/0608041].

\bibitem{R-e6-orhan-metin}O. Cakir and M. Yilmaz, Europhys. Lett.
\textbf{38,} \emph{13} (1997).

\bibitem{DSB}Y. Hosotani, Phys. Lett. B \textbf{126}, \emph{309}
(1983) ; B. McInnes, J. Math. Phys. \textbf{31}, \emph{2094} (1990). 

\bibitem{R-calchep}A. Pukhov, {[}arXiv:hep-ph/0412191]; E. Boos et
al. {[}CompHEP Collaboration], Nucl. Instrum. Meth. A \textbf{534},
\emph{250} (2004).

\bibitem{R-cteq}J. Pumplin, D.R. Stump, J. Huston, H.L. Lai, P. Nadolsky
and W.K. Tung, JHEP \textbf{0207}, 012 (2002) {[}arXiv:hep-ph/0201195].

\bibitem{Cpyth}A.S.Belyaev et al, {[}arXiv:hep-ph/0101232]

\bibitem{Pythia}T. Sjostrand et al., Computer Phys. Commun. \textbf{135}
(2001) 238 (LU TP 00-30, {[}arXiv:hep-ph/0010017])

\bibitem{ATLFast}E. Richter-Was et al., ATLAS Note PHYS-98-131(1998);
http://www.hep.ucl.ac.uk/atlas/atlfast/.

\bibitem{MadGraph}T. Stelzer and W. F. Long, Phys. Commun. \textbf{81},
357 (1994).
\end{thebibliography}
\end{document}